\documentclass[10pt,nofootinbib,twocolumn,aps,prd]{revtex4-1}
\usepackage{amsmath,mathptmx,amssymb}
\usepackage{graphicx,graphics,subfigure}
\usepackage{tabularx,ctable}
\usepackage[colorlinks=true,linkcolor=blue,citecolor=blue,urlcolor=blue]{hyperref}

\def\be{\begin{equation}}
\def\ee{\end{equation}}
\def\bea{\begin{eqnarray}}
\def\eea{\end{eqnarray}}
\def\ba{\begin{aligned}}
\def\ea{\end{aligned}}
\def\nn{\nonumber}
\def\p{\partial}

\begin{document}


\title{Topological classes of thermodynamics of the rotating charged AdS black holes in gauged supergravities}

\author{Xiao-Dan Zhu$^{1}$}
\email{zxdcwnu@163.com}

\author{Di Wu$^{1}$}
\email{Corresponding author: wdcwnu@163.com}

\author{Dan Wen$^{2}$}
\email{Corresponding author: wendan@cqupt.edu.cn}

\affiliation{$^{1}$School of Physics and Astronomy, China West Normal University,
Nanchong, Sichuan 637002, People's Republic of China \\
$^{2}$College of Science, Chongqing University of Posts and Telecommunications,
Nanan, Chongqing 400065, People's Republic of China}

\date{\today}

\begin{abstract}
In this paper, we investigate the topological numbers of rotating charged AdS black holes in both four- and five-dimensional gauged supergravity theories. Our analysis is conducted within the framework of the thermodynamical topological approach to black holes, utilizing the generalized off-shell Helmholtz free energy. We demonstrate that the number of rotation parameters plays a significant role in determining the topological numbers of five-dimensional rotating AdS black holes. Moreover, our findings indicate that the topological numbers of both four- and five-dimensional rotating AdS black holes are not influenced by the number of electric charge parameters. This highlights a distinct difference in how rotation and electric charge parameters impact the thermodynamic topological properties of these black holes.
\end{abstract}

\maketitle

\section{Introduction}
There has been a great deal of interest in supergravity theories as one of the possible ways to quantize gravity. In particular, with the development of the
remarkable anti-de Sitter/conformal field theory (AdS/CFT) correspondence \cite{ATMP2-231,PLB428-105,ATMP2-253}, it has become of considerable interest to investigate the thermodynamical properties of the rotating charged AdS black holes in four- and five-dimensional gauged supergravity theories \cite{PRD72-084028,
PRD84-024037,NPB717-246,PRL95-161301,hep-th/0504080,PRD72-041901,PRD73-104036,PRL100-121301,PRD80-044037,PRD80-084009,PRD83-044028,PRD83-121502,PLB707-286,
PLB726-404,PLB746-276,JHEP1121031,PRD101-024057,PRD102-044007,PRD103-044014}. In fact, deriving thermodynamic mass formulas \cite{PRD7-2333,PRD13-191,PRL30-71,
PRL25-1596,PRD4-3552,PRD100-101501,PRD105-124013,PRD108-064034,PRD108-064035,PLB846-138227} and exploring the chemistry of black holes \cite{PRD92-124069,CQG34-063001} are just part of the broader investigation into black hole thermodynamics.

Recently, topology has become a new effective tool to study the thermodynamical properties of black holes \cite{PRD105-104003,PRD105-104053,PLB835-137591,
PRD107-046013,PRD107-106009,JHEP0623115,2305.05595,2305.05916,2305.15674,2305.15910,2306.16117,PRD106-064059,PRD107-044026,PRD107-064015,2212.04341,2302.06201,
2304.14988,2309.00224,2312.12784,2402.18791,2403.14730,2404.02526}.\footnote{Topology can also be utilized to investigate the light rings \cite{PRL119-251102,
PRL124-181101,PRD102-064039,PRD103-104031,PRD105-024049,PRD108-104041,2401.05495} and the timelike circular orbits \cite{PRD107-064006,JCAP0723049,2406.13270}.} Remarkably, in Ref. \cite{PRL129-191101}, by considering black hole solutions as the thermodynamic topological defects, Wei \textit{et al.} proposed a new systematic approach for black hole classification according to the different topological numbers of black holes. Because of its simplicity as well as adaptability, the topological method provided in Ref. \cite{PRL129-191101} quickly developed favor and was successfully utilized to explore the topological numbers for many black holes \cite{PRD107-064023,JHEP0123102,PRD107-024024,PRD107-084002,PRD107-084053,2303.06814,2303.13105,2304.02889,2306.13286,2304.05695,2306.05692,2306.11212,
EPJC83-365,2306.02324,PRD108-084041,2307.12873,2309.14069,AP458-169486,2310.09602,2310.09907,2310.15182,2311.04050,2311.11606,2312.04325,2312.06324,2312.13577,
2312.12814,PS99-025003,2401.16756,AP463-169617,PDU44-101437,2403.14167,2404.08243,2405.02328,2405.07525,2405.20022,2406.08793,AC48-100853,2402.00106}. Very recently, we explored the topological numbers of static multi-charge AdS black holes within four- and five-dimensional gauged supergravities. We identified a novel temperature-dependent thermodynamic topological phase transition that occurs in the four-dimensional static-charged AdS black hole in Einstein-Maxwell-dilaton-axion gauged supergravity theory, the four-dimensional static-charged AdS Horowitz-Sen black hole, and the five-dimensional static-charged AdS black hole in Kaluza-Klein gauged supergravity theory \cite{2402.00106}. However, the topological numbers of rotating charged AdS black holes in four- and five-dimensional gauged supergravities, and whether similar temperature-dependent thermodynamic topological phase transitions occur, are still unclear and require further investigation. This is the motivation behind our current study.

In this paper, we will explore the effect of the number of charge parameters on the topological number of rotating AdS black holes in four- and five-dimensional gauged supergravity theory and the effect of the number of rotation parameters on the topological number of charged AdS black holes in five-dimensional gauged supergravity theory. We show that the topological numbers of the five-dimensional rotating AdS black holes are significantly influenced by the number of rotation parameters. Moreover, we show that the topological numbers of four- and five-dimensional rotating AdS black holes do not depend on the number of electric charge parameters.

The rest of this paper is organized as follows. In Sec. \ref{II}, we give a brief introduction to the thermodynamic topological approach proposed in Ref. \cite{PRL129-191101}. In Sec. \ref{III}, we study the topological numbers of four-dimensional rotating pairwise-equal four-charge AdS black holes in gauged supergravity theory \cite{NPB717-246} for two different combinations of electric charge parameters, respectively. In Sec. \ref{IV}, we investigate the topological numbers of five-dimensional rotating charged AdS black holes in minimal gauged supergravity theory \cite{PRL95-161301} by considering the singly-, and double-rotating cases, respectively. Furthermore, to facilitate the comparison of the effect of the charge parameter on the topological number, we also discuss the topological number of the five-dimensional double-rotating Kerr-AdS black hole. This paper ends with conclusions in Sec. \ref{V}.

\section{A brief introduction to the thermodynamic topological approach}\label{II}
In this section, we give a brief introduction to the novel thermodynamic topological approach proposed in Ref. \cite{PRL129-191101}. In the beginning, according to Ref. \cite{PRL129-191101}, one can give the generalized off-shell Helmholtz free energy as
\be\label{FE}
\mathcal{F} = M -\frac{S}{\tau}
\ee
for the black hole thermodynamic system with mass $M$ and entropy $S$, the extra variable $\tau$ can be regarded as the inverse temperature of the cavity surrounding the black hole. Only when $\tau = T^{-1}$, the generalized Helmholtz free energy manifests on-shell characteristics and returns to the standard Helmholtz free energy $F = M -TS$ of the black hole \cite{PRD15-2752,PRD33-2092,PRD105-084030,PRD106-106015}.

In Ref. \cite{PRL129-191101}, the key vector $\phi$ is given by
\bea\label{vector}
\phi = \Big(\frac{\p \mathcal{F}}{\p r_h}\, , ~ -\cot\Theta\csc\Theta\Big) \, ,
\eea
where $r_h$ is the event horizon radius of the black hole, and $0 \le \Theta \le \pi$. The component $\phi^\Theta$ diverges at $\Theta = 0$ and $\Theta = \pi$, and the direction of the vector points outward there.

One can apply Duan's $\phi$-mapping topological currents theory \cite{SS9-1072,NPB514-705,PRD61-045004} to establish the topological current as follows:
\be\label{jmu}
j^{\mu}=\frac{1}{2\pi}\epsilon^{\mu\nu\rho}\epsilon_{ab}\p_{\nu}n^{a}\p_{\rho}n^{b}\, , \qquad
\mu,\nu,\rho=0,1,2,
\ee
where $\p_{\nu}= \p/\p x^{\nu}$, $x^{\nu}=(\tau,~r_h,~\Theta)$. The unit vector is given as $n = (n^r, n^\Theta)$ with $n^r = \phi^{r_h}/||\phi||$ and $n^\Theta = \phi^{\Theta}/||\phi||$. It is easy to prove that the above topological current $j^\mu$ is conserved, i.e., $\p_{\mu}j^{\mu} = 0$. Employing the Jacobi tensor $\epsilon^{ab}J^{\mu}(\phi/x) = \epsilon^{\mu\nu\rho}\p_{\nu}\phi^a\p_{\rho}\phi^b$, the topological current could be expressed as a $\delta$-function of the field configuration
\be
j^{\mu}=\delta^{2}(\phi)J^{\mu}\Big(\frac{\phi}{x}\Big) \, .
\ee
This indicates that $j^\mu$ is nonzero only at the zero points of $\phi^a(x_i)$, namely, $\phi^a(x_i) = 0$. Consequently, one can employ the following formula to determine the topological number in the given parameter region $\Sigma$:
\be
W = \int_{\Sigma}j^{0}d^2x = \sum_{i=1}^{N}\beta_{i}\eta_{i} = \sum_{i=1}^{N}w_{i}\, ,
\ee
where the positive Hopf index $\beta_i$ represents the number of loops caused by $\phi^a$ in the vector $\phi$-space as $x^{\mu}$ moves surrounding the zero point $z_i$, the Brouwer degree $\eta_{i}= \mathrm{sign}(J^{0}({\phi}/{x})_{z_i})=\pm 1$, and $w_i$ is the winding number for the $i$th zero point of $\phi$.
Notably, one important tool for assessing local thermodynamic stability is the local winding number $w_{i}$. The thermodynamic stability of black holes is shown by positive $w_{i}$ values, while unstable ones are indicated by negative values. In a exact black hole solution at a given temperature, thermodynamically stable and unstable black holes are distinguished by the global topological number $W$.

\section{Four-dimensional rotating pairwise-equal four-charge AdS black holes in gauged supergravity theory}\label{III}
In this section, we will investigate the topological numbers of the four-dimensional rotating pairwise-equal four-charge AdS black holes in gauged supergravity theory \cite{NPB717-246}. The metric and two Abelian gauge potentials of this black hole are given by
\bea
ds_4^2 &=& -\frac{\Delta_r}{W}\left(dt -\frac{a\sin^2\theta}{\Xi}d\varphi \right)^2 +W\left( \frac{dr^2}{\Delta_r} +\frac{d\theta^2}{\Delta_\theta}\right) \nn \\ &&+\frac{\Delta_\theta\sin^2\theta}{W}\left(adt -\frac{r_1r_2 +a^2}{\Xi}d\varphi \right)^2 \, , \\
A_1 &=& \frac{2r_1\sqrt{2q_1(q_1 +m)}}{W}\left(dt -\frac{a\sin^2\theta}{\Xi}d\varphi \right) \, , \nn \\
A_2 &=& \frac{2r_2\sqrt{2q_2(q_2 +m)}}{W}\left(dt -\frac{a\sin^2\theta}{\Xi}d\varphi \right) \, , \nn
\eea
where
\bea
r_1 &=& r +2q_1 \, , \qquad r_2 = r +2q_2 \, , \nn \\
\Delta_r &=& r^2 +a^2 -2mr +\frac{r_1r_2}{l^2}(r_1r_2 +a^2) \, , \quad \Xi = 1 -\frac{a^2}{l^2} \, , \nn \\
\Delta_\theta &=& 1 -\frac{a^2}{l^2}\cos^2\theta \, , \qquad W = r_1r_2 +a^2\cos^2\theta \, ,  \nn
\eea
in which $a$ is the rotation parameter, $l$ is the AdS radius, $m$, $q_1$, and $q_2$ are the mass and two independent electric charge parameters, respectively.

In addition, there is a special truncated supergravity solution: when the electric charge parameters $q_1 \ne 0$ and $q_2 = 0$, this is known as the rotating charged AdS black hole solution in Einstein-Maxwell-dilaton-axion (EMDA) supergravity theory.

The thermodynamic quantities are given by \cite{PRD84-024037}
\be\ba\label{therm}
&M = \frac{m +q_1 +q_2}{\Xi^2} \, ,  \qquad S = \frac{\pi(r_1r_2 +a^2)}{\Xi} \, , \\
&Q_1 = Q_2 = \frac{\sqrt{q_1(q_1 +m)}}{2\Xi} \, , ~~ Q_3 = Q_4 = \frac{\sqrt{q_2(q_2 +m)}}{2\Xi} \, , \\
&T = \frac{\p_{r_h}\Delta(r_h)}{4\pi(r_1r_2 +a^2)} \, , \qquad \Omega = \frac{a(l^2 +r_1r_2)}{l^2(r_1r_2 +a^2)} \, ,  \\
&\Phi_1 = \Phi_2 = \frac{2r_1\sqrt{q_1(q_1 +m)}}{r_1r_2 +a^2} \, , \quad P = \frac{3}{8\pi l^2} \, , \\
&\Phi_3 = \Phi_4 = \frac{2r_2\sqrt{q_2(q_2 +m)}}{r_1r_2 +a^2} \, , \quad J = Ma \, , \\
&V = \frac{4\pi}{3\Xi}(r_h +q_1 +q_2)(r_1r_2 +a^2) +\frac{4\pi}{3}aJ \, ,
\ea\ee
(evaluated at $r = r_h$).

It is simple to prove that the above thermodynamic quantities (\ref{therm}) satisfy the first law and the Bekenstein-Smarr mass formula simultaneously
\bea
dM &=& TdS +\Omega dJ +\sum_{i=1}^4\Phi_idQ_i +VdP \, , \\
M &=& 2TS +2\Omega J +\sum_{i=1}^4\Phi_iQ_i -2VP \, .
\eea

Using the definition of the generalized off-shell Helmholtz free energy (\ref{FE}) and substituting the relation $l^2 = 3/(8\pi P)$ \cite{PRD84-024037,CPL23-1096,CQG26-195011}, one can arrive at
\bea
\mathcal{F} &=& \frac{4\pi P}{3r_h\Xi^2}(r_h +2q_1)(r_h +2q_2)\Big[r_h^2 +2r_h(q_1 +q_2) \nn \\
&&+4q_1q_2 +a^2 \Big] +\frac{1}{2r_h\Xi^2}\Big[r_h^2 +2r_h(q_1 +q_2) +a^2 \Big] \nn \\
&&-\frac{\pi}{\Xi\tau}\Big[r_h^2 +2r_h(q_1 +q_2) +4q_1q_2 +a^2 \Big]
\eea
for the rotating pairwise-equal four-charge AdS black hole in four-dimensional gauged supergravity theory. Then the components of the vector $\phi$ can be computed as
\bea
\phi^{r_h} &=& \frac{1}{2(3 -8\pi Pa^2)^2r_h^2\tau}\bigg\{12\pi r_h^2(8\pi Pa^2 -3)(r_h +q_1 +q_2) \nn \\
&&+3\Big\{3r_h^2 +8\pi P(r_h +2q_1)(r_h +2q_2)\big[3r_h^2 +2r_h(q_1 +q_2)  \nn \\
&&-4q_1q_2\big] +a^2\big[8\pi P(r_h^2 -4q_1q_2) -3\big]\Big\}\tau\bigg\} \, ,  \\
\phi^{\Theta} &=& -\cot\Theta\csc\Theta \, .
\eea
By solving the equation: $\phi^{r_h} = 0$, one can calculate the zero point of the vector field $\phi^{r_h}$ as
\begin{widetext}
\be\label{tau4d}
\tau = -\frac{4\pi r_h^2(8\pi Pa^2 -3)(r_h +q_1 +q_2)}{3r_h^2 +8\pi P(r_h +2q_1)(r_h +2q_2)[3r_h^2 +2r_h(q_1 +q_2) -4q_1q_2] +a^2[8\pi P(r_h^2 -4q_1q_2) -3]} \, .
\ee
\end{widetext}
It is important to note that Eq. (\ref{tau4d}) consistently reduces to the result that is derived from the case of the four-dimensional Kerr-AdS black hole \cite{PRD107-084002} when the two independent electric charge parameters $q_1$ and $q_2$ are zero. Due to the fact that there are two pairwise-equal electric charge parameters to consider, it will be much simpler to consider the topological numbers of the special two- and the general four-charge rotating AdS cases, respectively, as well as to compare the impact of the number of electric charge parameters on the topological number.

\subsection{Rotating charged AdS black hole in EMDA gauged supergravity}\label{IIIA}

\begin{figure}[b]
\centering
\includegraphics[width=0.35\textwidth]{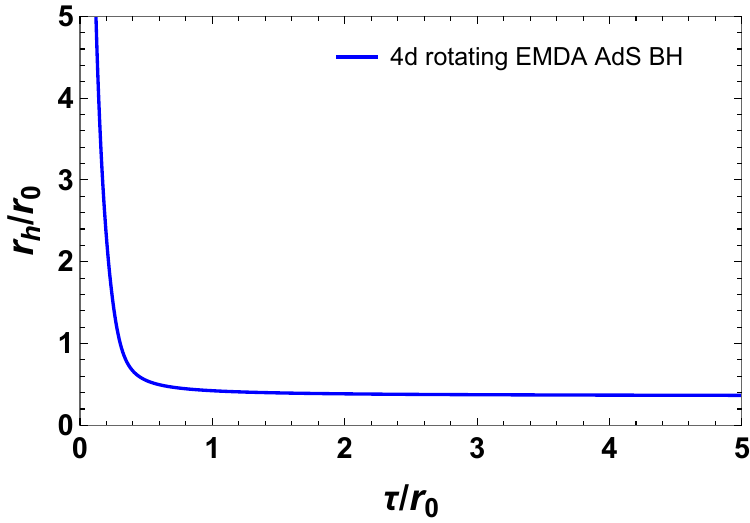}
\caption{Zero points of the vector $\phi^{r_h}$ shown in the $r_h-\tau$ plane with $q_1/r_0 = 1$, $Pr_0^2 = 0.1$, $a/r_0 = 1$, and $q_2 = 0$. There is one thermodynamically stable four-dimensional rotating charged AdS black hole in EMDA gauged supergravity theory for any value of $\tau$. Obviously, the topological number is: $W = 1$.
\label{4d2cBH}}
\end{figure}

\begin{figure}[b]
\centering
\includegraphics[width=0.35\textwidth]{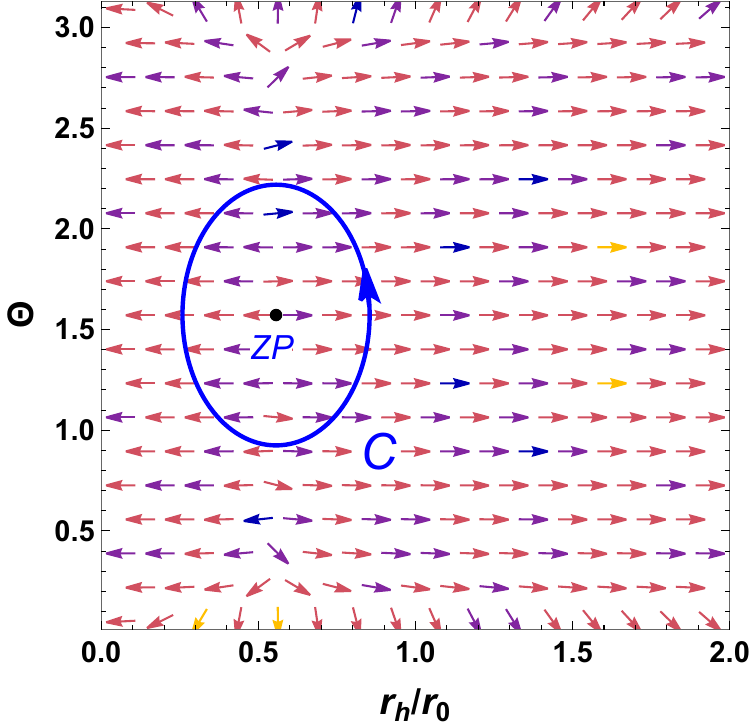}
\caption{The arrows represent the unit vector field $n$ on a portion of the $r_h-\Theta$ plane for the four-dimensional rotating charged AdS black hole in EMDA gauged supergravity theory with $\tau/r_0 = 0.5$, $q_1/r_0 = 1$, $Pr_0^2 = 0.1$, $a/r_0 = 1$, and $q_2 = 0$. The zero point (ZP) marked with a black dot is at $(r_h/r_0, \Theta) = (0.55,\pi/2)$. The blue contour $C$ is a closed loop enclosing the zero point.
\label{d42cBH}}
\end{figure}

For the four-dimensional rotating charged AdS black hole in EMDA gauged supergravity theory, one can plot the zero points of the component $\phi^{r_h}$ with $q_1/r_0 = 1$, $Pr_0^2 = 0.1$, $a/r_0 = 1$, and $q_2 = 0$ in Fig. \ref{4d2cBH}, and the unit vector field $n$ on a portion of the $\Theta-r_h$ plane in Fig. \ref{d42cBH} with $\tau/r_0 = 0.5$, respectively, where $r_0$ denotes an arbitrary length scale defined by the size of a cavity enclose the above black hole. As shown in Fig. \ref{4d2cBH}, there is only one thermodynamically stable four-dimensional rotating charged AdS black hole in EMDA gauged supergravity theory for any value of $\tau$. In Fig. \ref{d42cBH}, one can observe that there is a zero point located at $(r_h/r_0, \Theta) = (0.55,\pi/2)$, thus the winding number for the blue contour $C$ is $w = 1$. Based upon the local property of the zero point, one can easily obtain the topological number $W = 1$ for the four-dimensional rotating charged AdS black hole in EMDA gauged supergravity theory, which is the same as that of the four-dimensional Kerr-AdS and Kerr-Newman-AdS black holes \cite{PRD107-084002}.

\subsection{Rotating pairwise-equal four-charge AdS case}\label{IIIB}

\begin{figure}[t]
\centering
\includegraphics[width=0.35\textwidth]{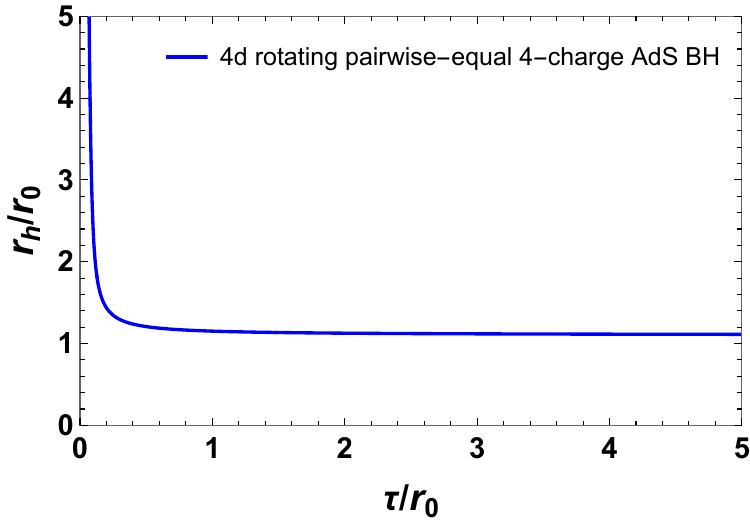}
\caption{Zero points of the vector $\phi^{r_h}$ shown in the $r_h-\tau$ plane with $q_1/r_0 = 1$, $q_2/r_0 = 3$, $Pr_0^2 = 0.1$, and $a/r_0 = 1$. There is one thermodynamically stable four-dimensional rotating pairwise-equal four-charge AdS black hole in gauged supergravity theory for any value of $\tau$. Obviously, the topological number is: $W = 1$.
\label{4d4cBH}}
\end{figure}

\begin{figure}[t]
\centering
\includegraphics[width=0.35\textwidth]{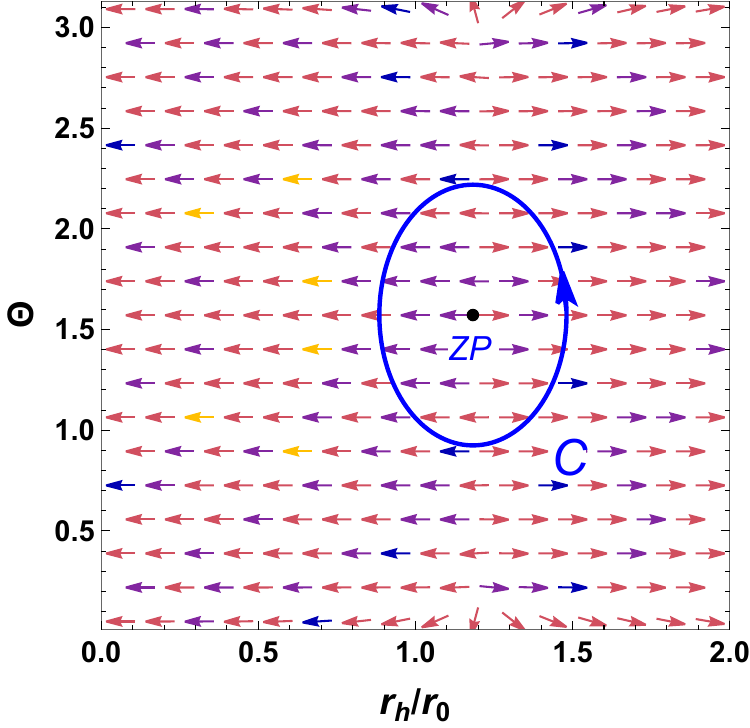}
\caption{The arrows represent the unit vector field $n$ on a portion of the $r_h-\Theta$ plane for the four-dimensional rotating pairwise-equal four-charge AdS black hole in gauged supergravity theory with $\tau/r_0 = 0.5$, $q_1/r_0 = 1$, $q_2/r_0 = 3$, $Pr_0^2 = 0.1$, and $a/r_0 = 1$. The zero point (ZP) marked with a black dot is at $(r_h/r_0, \Theta) = (1.21,\pi/2)$. The blue contour $C$ is a closed loop enclosing the zero point.
\label{d44cBH}}
\end{figure}

Taking $q_1/r_0 = 1$, $q_2/r_0 = 3$, $Pr_0^2 = 0.1$, and $a/r_0 = 1$ for the four-dimensional general rotating pairwise-equal four-charge AdS black hole black hole in gauged supergravity theory, we plot zero points of $\phi^{r_h}$ in the $r_h-\tau$ plane in Fig. \ref{4d4cBH}, and the unit vector field $n$ on a portion of the $\Theta-r_h$ plane with $\tau/r_0 = 0.5$ in Fig. \ref{d44cBH}, respectively. Obviously, there is only one thermodynamically stable four-dimensional rotating pairwise-equal four-charge AdS black hole in gauged supergravity theory for any value of $\tau$. It is easy to observe that the zero point is located at $(r_h/r_0, \Theta) = (1.21,\pi/2)$ from Fig. \ref{d44cBH}. Based upon the local property of the zero point, we can easily obtain the topological number $W = 1$ for the four-dimensional rotating pairwise-equal four-charge AdS black hole, which is also the same as that of the four-dimensional Kerr-AdS and Kerr-Newman-AdS black holes \cite{PRD107-084002}, and the four-dimensional rotating charged AdS black hole in EMDA gauged supergravity theory in the last subsection.

\subsection{Discussions}
In this subsection, we explore the underlying reasons why multiple electric charge parameters do not alter the topological number of four-dimensional rotating AdS black holes. Now, according to Ref. \cite{2402.00106}, one can construct a new vector
\be\label{Psi}
\Psi^{r_h} = \frac{\p\tau}{\p r_h} \, , \qquad \Psi^\Theta = -\cot\Theta\csc\Theta \, .
\ee
Then, by solving the equation $\Psi^{r_h} = 0$ and taking the limit $r \to 0$, one can obtain the following critical relationship:
\be
P_c = \frac{3}{8\pi a^2} \, .
\ee
Therefore, the presence or absence of generation and annihilation points is independent of the electric charge parameters. Additionally, according to the results given in our previous work \cite{PRD107-084002}, the generation and annihilation points of a four-dimensional Kerr-AdS black hole appear or disappear simultaneously in pairs, thereby not changing its topological number. These conclusions also hold for rotating charged AdS black holes in four-dimensional gauged supergravity theories.

\medskip

\section{Five-dimensional rotating charged AdS black holes in minimal gauged supergravity theory}\label{IV}
In this section, we would like to derive the topological numbers of the five-dimensional rotating charged AdS black holes in minimal gauged supergravity theory \cite{PRL95-161301}. The line element of this black hole can be recast into a simple form in terms of the Boyer-Lindquist coordinates as follows \cite{PRD80-084009}:
\bea\label{5dsolution}
ds_5^2 &=& -\frac{\Delta_r}{\Sigma}X^2 +\frac{\Sigma}{\Delta_r}dr^2 +\frac{\Sigma}{\Delta_\theta}d\theta^2 +\left(\frac{ab}{r\rho}Z +\frac{q\rho}{r\Sigma}X \right)^2 \nn \\
&&+\frac{\Delta_\theta(a^2 -b^2)^2\sin^2\theta\cos^2\theta}{\rho^2\Sigma}Y^2 \, ,
\eea
and the Abelian gauge potential is
\be
A = \frac{\sqrt{3}q}{\Sigma}X \, ,
\ee
where we denote
\be\ba
&X = dt -\frac{a\sin^2\theta}{\Xi_a}d\varphi -\frac{b\cos^2\theta}{\Xi_b}d\psi \, , \\
&Y = dt -\frac{(r^2 +a^2)a}{(a^2 -b^2)\Xi_a}d\varphi -\frac{(r^2 +b^2)b}{(b^2 -a^2)\Xi_b}d\psi \, , \\
&Z = dt -\frac{(r^2 +a^2)\sin^2\theta}{a\Xi_a}d\varphi -\frac{(r^2 +b^2)\cos^2\theta}{b\Xi_b}d\psi \, ,
\ea\ee
and
\bea
\Delta_r &=& (r^2 +a^2)(r^2 +b^2)\left(\frac{1}{r^2} +\frac{1}{l^2} \right) -2m +\frac{q^2 +2qab}{r^2} \, , \nn \\
\Delta_\theta &=& 1 -\frac{\rho^2}{l^2} \, , \quad \Sigma = r^2 +\rho^2 \, , \quad \Xi_a = 1 -\frac{a^2}{l^2} \, , \nn \\
\rho &=& \sqrt{a^2\cos^2\theta +b^2\sin^2\theta} \, , \qquad \Xi_b = 1 -\frac{b^2}{l^2} \, . \nn
\eea
Here the parameters ($m$, $q$, $a$, $b$, $l$) are the mass parameter, the electric charge parameter, and two independent rotation parameters of the black hole as well as the AdS radius. We point out that this solution (\ref{5dsolution}) consistently reduces to the  five-dimensional double-rotating Kerr-AdS black hole case \cite{AP172-304} when the electric charge parameter $q$ is turned off.

The above black hole has the thermodynamic quantities \cite{PRD84-024037}
\be\ba\label{Therm}
&M = \frac{\pi ml^2(2\Xi_a +2\Xi_b -\Xi_a\Xi_b) +2\pi qab(\Xi_a +\Xi_b)}{4\Xi_a^2\Xi_b^2l^2} \, , \ \\
&S = \frac{\pi^2[(r_h^2 +a^2)(r_h^2 +b^2) +abq]}{2\Xi_a\Xi_br_h} \, ,   \\
&J_a = \frac{\pi[2ma +qb(2 -\Xi_a)]}{4\Xi_a^2\Xi_b} \, , \quad Q = \frac{\sqrt{3}\pi q}{4\Xi_a\Xi_b} \, , \quad \ \\
&J_b = \frac{\pi[2mb +qa(2 -\Xi_b)]}{4\Xi_a\Xi_b^2} \, , \quad P = \frac{3}{4\pi l^2} \, , \quad  \\
&T = \frac{r_h^4(l^2 +r_h^2 +a^2 +b^2) -(ab +q)^2l^2}{2\pi r_hl^2[(r_h^2 +a^2)(r_h^2 +b^2) +abq]} \, ,  \\
&\Phi = \frac{\sqrt{3}qr_h^2}{(r_h^2 +a^2)(r_h^2 +b^2) +abq} \, ,  \\
&\Omega_a = \frac{a(r_h^2 +b^2)(l^2 +r_h^2) +bql^2}{l^2[(r_h^2 +a^2)(r_h^2 +b^2) +abq]} \, ,  \\
&\Omega_b = \frac{b(r_h^2 +a^2)(l^2 +r_h^2) +aql^2}{l^2[(r_h^2 +a^2)(r_h^2 +b^2) +abq]} \, ,  \\
&V = \frac{\pi^2}{6\Xi_a\Xi_b}[3(r_h^2 +a^2)(r_h^2 +b^2) +2abq]   \\
&\qquad+\frac{2\pi}{3}(aJ_a +bJ_b) \, .
\ea\ee

One can easily verify that the above thermodynamic quantities completely obey the first law and the Bekenstein-Smarr mass formula simultaneously,
\be
dM = TdS +\Omega_adJ_a +\Omega_bdJ_b +\Phi dQ +VdP \, ,
\ee
\be
2M = 3(TS +\Omega_aJ_a +\Omega_bJ_b) +2(\Phi Q -VP) \, .
\ee

Utilizing Eq. (\ref{Therm}) and and substituting the relation $l^2 = 3/(4\pi P)$, one can calculate the generalized Helmholtz free energy as
\bea
\mathcal{F} &=& \frac{\pi}{24\Xi_a^2\Xi_b^2}\bigg\{\frac{3}{r_h^2}\Big[(r_h^2 +a^2)(r_h^2 +b^2)\Big(\frac{4\pi P}{3}r_h^2 +1\Big) \nn \\
&& +q^2 +2abq\Big]\Big[2\Xi_a +2\Xi_b -\Xi_a\Xi_b \Big] +16\pi abqP \nn \\ &&-\frac{12\pi\Xi_a\Xi_b}{r_h\tau}\Big[(r_h^2 +a^2)(r_h^2 +b^2) +abq\Big]\bigg\} \, ,
\eea
then the components vector $\phi$ can be easily computed as
\bea
\phi^{r_h} &=& \pi(12r_h^3\Xi_a^2\Xi_b^2\tau)^{-1}\bigg\{\Big\{-r_h^4\Big[3 +4\pi P(a^2 +b^2)\Big] \nn \\
&&-8\pi Pr_h^6 +3(q +ab)^2  \Big\}\Big[\Xi_a\Xi_b -2(\Xi_a +\Xi_b)\Big]\tau \nn \\
&&-6\pi\Xi_a\Xi_br_h\Big[3r_h^4 +(a^2 +b^2)r_h^2 -ab(ab +q) \Big]\bigg\} \, ,\quad \nn \\
\phi^{\Theta} &=& -\cot\Theta\csc\Theta \, . \nn
\eea
By solving the equation: $\phi^{r_h} = 0$, and it is easy to obtain
\bea
\tau &=& 6\pi r_h(4\pi Pa^2 -3)(4\pi Pb^2 -3)\Big[ab(ab +q) -r_h^2(a^2 +b^2) \nn \\
&&-3r_h^4\Big]\bigg\{4\pi P\Big[a^2(4\pi Pb^2 +3) +3b^2\Big] -27\bigg\}^{-1}\bigg\{8\pi Pr_h^6 \nn \\
&&+\Big[4\pi P (a^2 +b^2) +3\Big]r_h^4 -3(ab +q)^2\bigg\}^{-1}
\eea
as the zero point of the vector field $\phi$, which consistently reduces to the one obtained in the five-dimensional singly-rotating Kerr-AdS black hole case
\cite{PRD107-084002} when the electric charge parameter $q$ and the second rotation parameter $b$ are turned off.

Similar to Sec. \ref{III}, it will be more convenient to consider the topological numbers of the singly- and double-rotating charged AdS black hole cases, respectively. By the way, to make it easier to see the effect of the charge parameter on the topological number, we also discuss the topological number of the neutral double-rotating AdS black hole case (i.e., the five-dimensional double-rotating Kerr-AdS black hole case) in Sec. \ref{IVC}.

\subsection{Singly-rotating charged AdS case}\label{IVA}

\begin{figure}[t]
\centering
\includegraphics[width=0.35\textwidth]{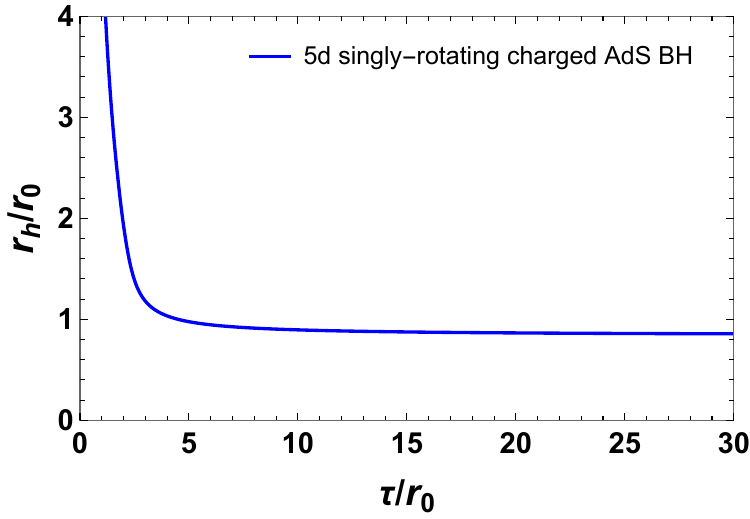}
\caption{Zero points of the vector $\phi^{r_h}$ shown in the $r_h-\tau$ plane with $a/r_0 = 1$, $q/r_0^2 = 1$, $b = 0$, and $Pr_0^2 = 0.1$. There is one thermodynamically stable five-dimensional singly-rotating charged AdS black hole in minimal gauged supergravity theory for any value of $\tau$. Obviously, the topological number is: $W = 1$.
\label{5d1rBH}}
\end{figure}

\begin{figure}[t]
\centering
\includegraphics[width=0.35\textwidth]{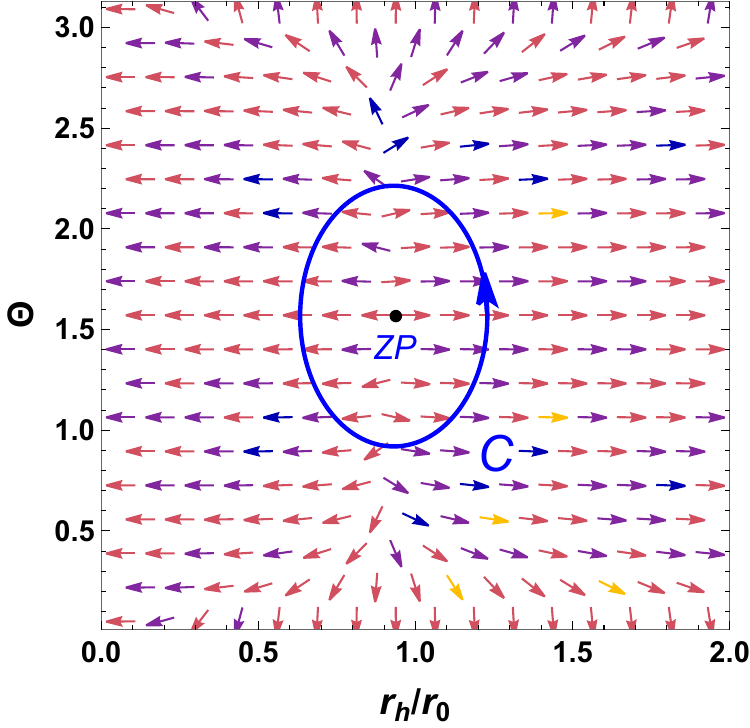}
\caption{The arrows represent the unit vector field $n$ on a portion of the $r_h-\Theta$ plane for the five-dimensional singly-rotating charged AdS black hole in minimal gauged supergravity theory with $\tau/r_0 = 10$, $a/r_0 = 1$, $q/r_0^2 = 1$, $b = 0$, and $Pr_0^2 = 0.1$. The zero point (ZP) marked with a black dot is at $(r_h/r_0, \Theta) = (0.90,\pi/2)$. The blue contour $C$ is a closed loop enclosing the zero point.
\label{d51rBH}}
\end{figure}

In Figs. \ref{5d1rBH} and \ref{d51rBH}, taking $a/r_0 = 1$, $q/r_0^2 = 1$, $b = 0$, and $Pr_0^2 = 0.1$ for the five-dimensional singly-rotating charged AdS black hole in minimal gauged supergravity theory, we plot the zero points of $\phi^{r_h}$ in the $r_h-\tau$ plane and the unit vector field $n$ with $\tau/r_0 = 10$, respectively. Note that for these values of $a/r_0$, $q/r_0^2$, and $Pr_0^2$, there is just one thermodynamically stable five-dimensional singly-rotating charged AdS black hole in minimal gauged supergravity theory for any value of $\tau$. In Fig. \ref{d51rBH}, one can find that the zero point is located at $(r_h/r_0, \Theta) = (0.90,\pi/2)$. As a result, the topological number $W = 1$ for the above black hole in five-dimensional minimal gauged supergravity theory can be explicitly established in Figs. \ref{5d1rBH} and \ref{d51rBH} via the local property of the zero point, which is identical to that of the five-dimensional singly-rotating Kerr-AdS black hole \cite{PRD107-084002}.

\subsection{Double-rotating charged AdS case}\label{IVB}

\begin{figure}[t]
\centering
\includegraphics[width=0.35\textwidth]{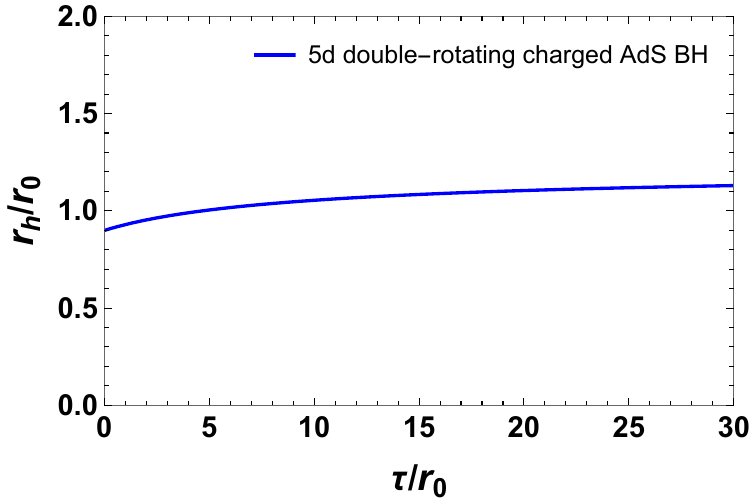}
\caption{Zero points of the vector $\phi^{r_h}$ shown in the $r_h-\tau$ plane with $a/r_0 = 1$, $b/r_0 = 2$, $q/r_0^2 = 1$, and $Pr_0^2 = 0.1$. There is one thermodynamically unstable five-dimensional double-rotating charged AdS black hole in minimal gauged supergravity theory for any value of $\tau$. Obviously, the topological number is: $W = -1$.
\label{5d2rBH}}
\end{figure}

\begin{figure}[t]
\centering
\includegraphics[width=0.35\textwidth]{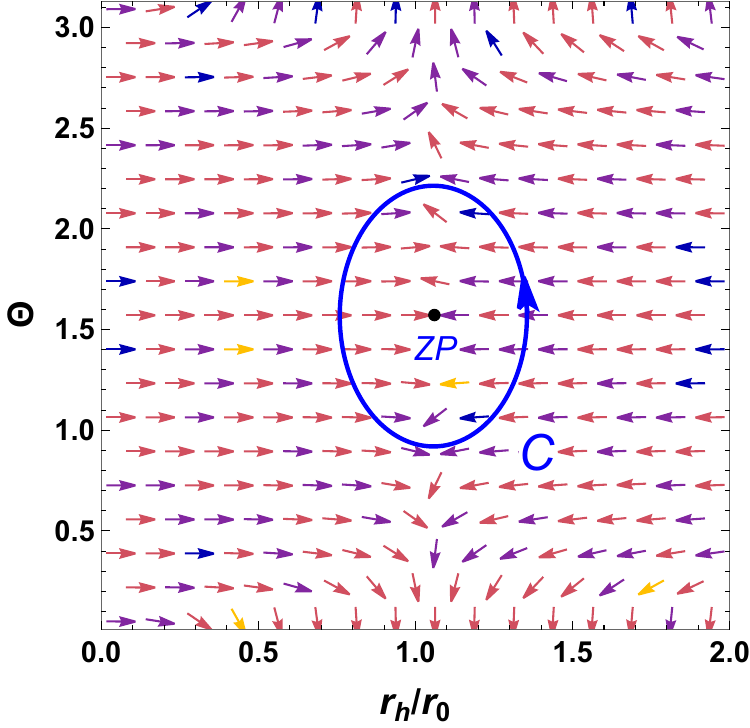}
\caption{The arrows represent the unit vector field $n$ on a portion of the $r_h-\Theta$ plane for the five-dimensional double-rotating charged AdS black hole in minimal gauged supergravity theory with $\tau/r_0 = 10$, $a/r_0 = 1$, $b/r_0 = 2$, $q/r_0^2 = 1$, and $Pr_0^2 = 0.1$. The zero point (ZP) marked with a black dot is at $(r_h/r_0, \Theta) = (1.05,\pi/2)$. The blue contour $C$ is a closed loop enclosing the zero point.
\label{d52rBH}}
\end{figure}

Considering the pressure $Pr_0^2 = 0.1$, the electric charge parameter $q/r_0^2 = 1$, two independent rotation parameters $a/r_0 = 1$ as well as $b/r_0 = 2$ for the five-dimensional double-rotating charged AdS black hole in minimal gauged supergravity theory, we plot the zero points of $\phi^{r_h}$ in the $r_h-\tau$ plane in Fig. \ref{5d2rBH}, and the unit vector field $n$ on a portion of the $\Theta-r_h$ plane with $\tau/r_0 = 10$ in Fig. \ref{d52rBH}. Indeed, there is only one thermodynamically unstable five-dimensional double-rotating charged AdS black hole in minimal gauged supergravity theory for whatever value of $\tau$. In Fig. \ref{d52rBH}, one can observe that the zero point is located at $(r_h/r_0, \Theta) = (1.05,\pi/2)$. Based on the local property of the zero point, we can easily obtain the topological number $W = -1$ for this black hole, which is different from that of the singly-rotating charged AdS case ($W = 1$) in the last subsection. This indicates that, in five-dimensional minimal gauged supergravity theory, the topological number for the rotating charged black holes is substantially influenced by the number of the rotation parameters.

\subsection{Double-rotating Kerr-AdS case}\label{IVC}

\begin{figure}[t]
\centering
\includegraphics[width=0.35\textwidth]{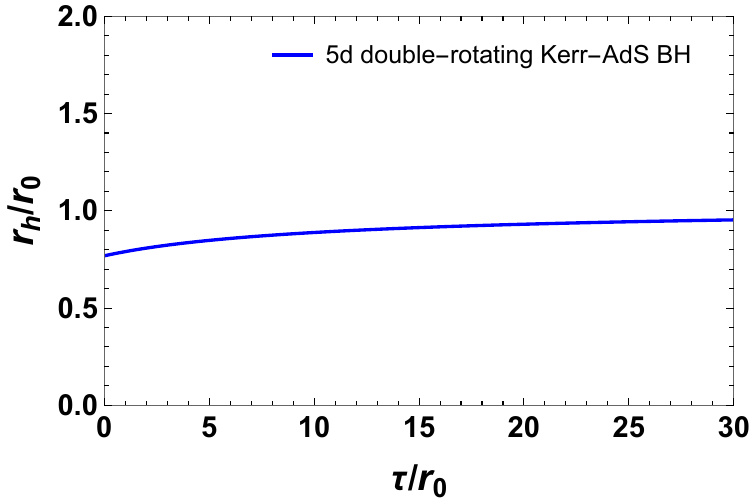}
\caption{Zero points of the vector $\phi^{r_h}$ shown in the $r_h-\tau$ plane with $a/r_0 = 1$, $b/r_0 = 2$, $q = 0$, and $Pr_0^2 = 0.1$. There is only one thermodynamically unstable five-dimensional double-rotating Kerr-AdS black hole for any value of $\tau$. Obviously, the topological number is: $W = -1$.
\label{5d2r0cBH}}
\end{figure}

\begin{figure}[t]
\centering
\includegraphics[width=0.35\textwidth]{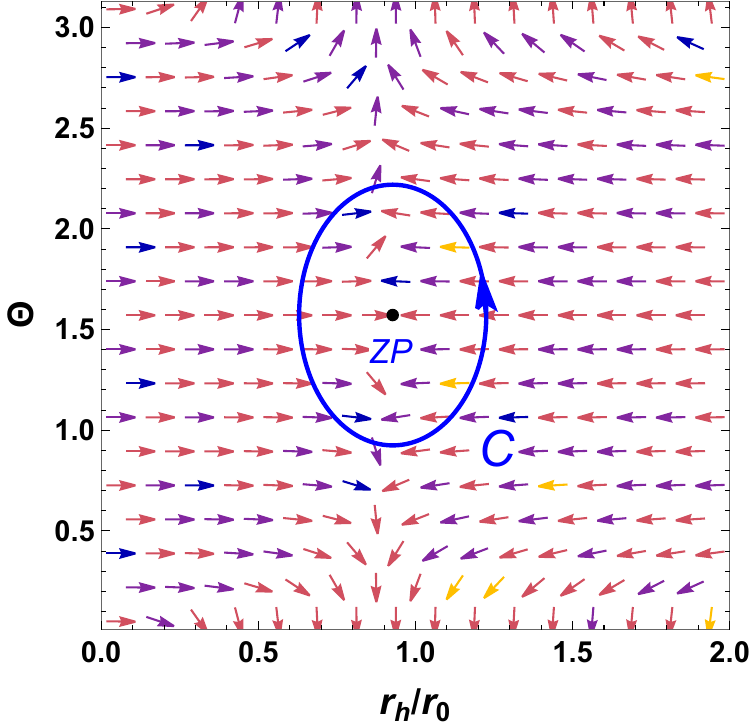}
\caption{The arrows represent the unit vector field $n$ on a portion of the $r_h-\Theta$ plane for the five-dimensional double-rotating Kerr-AdS black hole with $\tau/r_0 = 10$, $a/r_0 = 1$, $b/r_0 = 2$, $q = 0$, and $Pr_0^2 = 0.1$. The zero point (ZP) marked with a black dot is at $(r_h/r_0, \Theta) = (0.88,\pi/2)$. The blue contour $C$ is a closed loop enclosing the zero point.
\label{d52r0cBH}}
\end{figure}

In the following, in order to discuss the the effect of the charge parameter on the topological number, we consider the neutral double-rotating AdS black hole case, i.e., the five-dimensional double-rotating Kerr-AdS black hole case \cite{AP172-304}. Similar to the procedure adopted before, for the five-dimensional double-rotating Kerr-AdS black hole, we show the zero points of the component $\phi^{r_h}$ with $a/r_0 = 1$, $b/r_0 = 2$, $q = 0$, and $Pr_0^2 = 0.1$ in Fig. \ref{5d2r0cBH}, and the unit vector field $n$ on a portion of the $\Theta-r_h$ plane with $\tau/r_0 = 10$, $a/r_0 = 1$, $b/r_0 = 2$, $q = 0$, and $Pr_0^2 = 0.1$ in Fig. \ref{d52r0cBH}, respectively. There is clearly only one thermodynamically unstable five-dimensional double-rotating Kerr-AdS black hole for all values of $\tau$. The zero point can be observed at $(r_h/r_0, \Theta) = (0.88,\pi/2)$ in Fig. \ref{d52r0cBH}. Hence, utilizing the local property of the zero point, one can simply determine the topological number $W = -1$ for the five-dimensional double-rotating Kerr-AdS black hole, which is different from that of the five-dimensional singly-rotating Kerr-AdS black hole ($W = 1$) \cite{PRD107-084002}. In addition, one can demonstrate that the electric charge parameter has no effect on the topological number of five-dimensional rotating AdS black holes due to the fact that the topological number of the five-dimensional double-rotating AdS black hole in gauged supergravity theory in the last subsection is $W = -1$.

\subsection{Discussions}
In this subsection, we aim to investigate why multiple electric charge parameters do not affect the topological number of five-dimensional double-rotating AdS black holes. By utilizing the definition of the vector $\Psi$ in Eq. (\ref{Psi}) and solving the equation $\Psi^{r_h} = 0$, two critical pressure values are obtained as
\be
P_{c_1} = \frac{3}{4\pi a^2} \, , \qquad P_{c_2} = \frac{3}{4\pi b^2} \, .
\ee
Thus, the presence or absence of generation and annihilation points remains unaffected by the electric charge parameters. Moreover, as indicated in Ref. \cite{PRD107-084002}, the generation and annihilation points of the five-dimensional singly-rotating Kerr-AdS black hole emerge or vanish in pairs, and thus do not change its topological number. These findings are equally applicable to double-rotating charged AdS black holes in five-dimensional gauged supergravity theories.


\section{Conclusions}\label{V}

\begin{table}[b]
\caption{The topological number $W$, numbers of generation and annihilation points for the four- and five-dimensional rotating charged AdS black holes in gauged supergravity.}
\resizebox{0.48\textwidth}{!}{
\begin{tabular}{c|c|c|c}
\hline\hline
BH solution & $W$ & Generation point &Annihilation point\\ \hline
4d rotating EMDA AdS BH  & 1 & 0 & 0\\
4d rotating pairwise-equal four-charge AdS BH & 1 & 0 & 0\\
4d Kerr-AdS BH \cite{PRD107-084002} & 1 & 1 or 0 & 1or 0\\
4d Kerr-Newman-AdS BH \cite{PRD107-084002} & 1 & 1 or 0 & 1or 0\\ \hline
5d singly-rotating charged AdS BH & 1 & 0 & 0 \\
5d singly-rotating Kerr-AdS BH \cite{PRD107-084002} & 1 & 1 or 0 & 1or 0\\
5d double-rotating charged AdS BH & -1 & 0 & 0 \\
5d double-rotating Kerr-AdS BH & -1 & 0 & 0 \\
\hline\hline
\end{tabular}}
\label{TableI}
\end{table}

Combined ours with those in Ref. \cite{PRD107-084002}, Table \ref{TableI} summarizes some interesting results.

In this paper, using the generalized off-shell Helmholtz free energy, we investigate the topological number of the four-dimensional rotating charged AdS black holes in gauged supergravity theory \cite{NPB717-246} for two different combinations of electric charge parameters, respectively. We have also adopted the same method to discuss the topological number of the five-dimensional rotating charged AdS black holes in minimal gauged supergravity theory \cite{PRL95-161301} by considering the singly- and double-rotating cases, respectively. In addition, in order to explore the the impact of the charge parameter on the topological number, we also consider the five-dimensional double-rotating Kerr-AdS black hole case. We find that, in four-dimensional gauged supergravity theory, the topological numbers of the rotating pairwise-equal four-charge AdS black hole and the rotating charged AdS black hole in EMDA gauged supergravity have both $W = 1$. What is more, we demonstrate that, in five-dimensional minimal gauged supergravity theory, the topological number of the singly-rotating charged AdS black hole is $W = 1$, while that of the double-rotating charged AdS black hole is $W = -1$. In addition, one can indicate that the topological number of the five-dimensional double-rotating Kerr-AdS black hole is also $W = -1$.

In this work, we have obtained two particularly exciting consequences: (i) the number of rotation parameters has an important effect on the topological numbers of five-dimensional rotating AdS black holes, and (ii) the number of the electric charge parameters does not contribute to the topological numbers of the four- and five-dimensional rotating AdS black holes. A most related issue is to investigate the topological numbers of black holes in some modified gravity theories,
e.g., the scalar-tensor-vector gravity \cite{2406.00579}, Kalb-Ramond gravity \cite{2406.13461,2407.07416}, etc.

\acknowledgments
We are greatly indebted to the anonymous referee for his/her constructive comments to improve the presentation of this work. This work is supported by the National Natural Science Foundation of China (NSFC) under Grants No. 12205243, No. 12375053, No. 12205032, and No. 12275037, by the Sichuan Science and Technology Program under Grant No. 2023NSFSC1347, by the Program of Chongqing Human Resources and Social Security Bureau under Grant No. cx2021044, by the Doctoral Research Initiation Project of China West Normal University under Grant No. 21E028, and by the Talent Introduction Program of Chongqing University of Posts and Telecommunications under Grant No. E012A2021209.

\end{document}